# Single-crystalline gold microplates grown on substrates by solution-phase synthesis

*Xiaofei Wu[1,2], René Kullock[1], Enno Krauss[1], and Bert Hecht[1,]\**

[1] Nano-Optics and Biophotonics Group, Experimentelle Physik 5, Universität Würzburg, Am Hubland, 97074 Würzburg, Germany
[2] Present address: Experimentalphysik III, Universität Bayreuth, Universitätsstraße 30, 95447 Bayreuth, Germany
\* E-Mail: hecht@physik.uni-wuerzburg.de

**Abstract:** Chemically synthesized single-crystalline gold microplates have been attracting increasing interests because of their potential as high-quality gold films for nanotechnology. We present the growth of tens of nanometer thick and tens of micrometer large single-crystalline gold plates directly on solid substrates by solution-phase synthesis. Compared to microplates deposited on substrates from dispersion phase, substrate-grown plates exhibit significantly higher quality by avoiding severe small-particle contamination and aggregation. Substrate-grown gold plates also open new perspectives to study the growth mechanism via intermittent growth and observation cycles of a large number of individual plates. Growth models are proposed to interpret the evolution of thickness, area and shape of plates. It is found that the plate surface remains smooth after regrowth, implying the application of regrowth for producing giant plates as well as unique single-crystalline nano-structures.

## 1 Introduction

Gold films play an important role in various fields of nanotechnology due to their stability against corrosion, large electric conductivity, and the possibility to perform well-defined surface chemistry that can be used to build molecular architectures [1,2]. Thin gold films are also major materials for the fabrication of plasmonic nano-structures for various applications, such as optical antennas [3,4], plasmonic nano-circuits [5], surface-enhanced Raman scattering (SERS) [6], biological or chemical sensors [7], and photovoltaic devices [8].

Recently, the emergence of numerous synthesis approaches of plate-like gold crystals [9–18] has attracted the attention of researchers in different fields. Such gold plates are thought to result from seeds that exhibit at least two twin defects in the stacking of gold {111} planes, which lead to side facets consisting of {100} and {111} planes (Figure S1, Supporting Information). The existence of {100} planes accelerates the growth of side facets very much such that the plates grow much faster in lateral direction than in thickness [19–21]. Besides applications of gold nano-plates in SERS, sensing, catalysis, etc. [22], the mi-crometer-sized (lateral size in the order of 10 μm) plates are of particular interests, since they are large enough to act as gold films for nanotechnology [23–30]. The single-crystalline nature and ultra-smooth surface of gold plates lead to the unique properties of the systems and structures based on them. For instance, we previously reported plasmonic nano-structures fabricated from gold microplates using focused ion beam (FIB) milling [30]. These nano-structures exhibit significantly more precise shapes and favorable optical properties compared to those made from evaporated gold films, for which the random crystal grains degrade the properties of nano-structures. Moreover, compared to other methods of producing gold films of well-defined crystallinity [31–33], gold plate synthesis is simple, cheap and requires no specialized instrumentation. All these advantages imply that chemically synthesized gold microplates are promising candidates for high-quality gold films for nanotechnology.

In previous works, gold microplates were typically processed by drop-casting or spin-coating a dispersion of purified plates in solvent onto a substrate. However, as we show below, due to the poor quality caused by problems like small-particle contamination and aggregation, the use of gold microplates prepared in such ways is severely restricted.

Here, we report the growth of tens of nanometer thick and tens of micrometer large single-crystalline gold plates directly on solid substrates during solution-phase synthesis. Direct growth on substrates results in plates attaching to substrates perfectly and avoids aggregation and bending, and therefore exhibiting extraordinarily high quality. Such substrate-grown plates can be easily transferred to arbitrary substrates, in particular those that are not suitable for plate growth, without loss of quality. We also performed intermittent growth and observation cycles to study the time dependence of plate growth and the mechanism. The observed linear increase of the plate area with time suggests a sequential attachment of gold atoms to the plate side facets. Understanding the growth mechanism is crucial for manipulating the growth process, e.g. in order to adjust the plate thickness or to further increase the plate area. It is also found that the plate surface remains smooth after regrowth in solution.

## 2  Experimental Section

Gold microplates grown on substrates were achieved by simply immersing suitable substrates in a gold microplate synthesis solution. The synthesis follows the approach reported in [9] with some modification on the reaction conditions, e.g. the temperature and concentrations of reagents, and a glass weighing bottle with lid is used as the reaction container. The substrates are cleaned with ultrasonic bath of acetone and ethanol, without further treatment. They are then put into the container before adding the chemicals and kept vertical with help of a Teflon holder. After a certain synthesis period (normally 72 hours), a lot of microplates are found both on the substrates and in the solution (precipitated). The substrates are taken out of the solution, rinsed in ethanol to remove the chemicals, and dried with nitrogen blow. The rinsing and blowing do not have to be mild, as the plates attach to the surface quite tightly.

In order to gain insight into the plate growth behavior on substrate, we performed intermittent growth experiment on glass coverslips. After an initial growth period of 24 hours, the plates grown on coverslips were inspected using a transmission optical microscope and pictures were recorded to determine their lateral dimensions. The thickness of the plates in these pictures was also determined using a transmission spectrum method (see below). After these measurements, the coverslips were put back to the same solution for a regrowth period of 48 hours without further processing. After that, the same measurements were performed again, i.e. pictures of the same locations on the coverslips were recorded and the thickness of the same plates was measured.

To study the regrowth more carefully, intermittent growth experiment was also performed on a piece of $SiO_2$/Si wafer and plates were inspected with scanning electron microscopy (SEM) before and after regrowth (throughout this paper, "before regrowth" is used to denote the state after the initial growth period).

For the intermittent growth experiments, the substrates were rinsed in ethanol and dried with nitrogen for both sets of measurements before and after regrowth. During the measurements before regrowth, the synthesis solution was kept in oven at the synthesis temperature to be reused for the regrowth period. For the glass coverslip substrates, a scratch grid was made on the surface manually with a diamond scribe, while for the $SiO_2$/Si substrate, an array of a particular pattern was made on the surface with focused ion beam milling, in order to enable reliable identification of individual plates.

The area of a plate was obtained by multiplying the number of pixels occupied by the plate in the optical micrograph by the area per pixel. The plate thickness measurement was performed with a home-built setup consisting of the transmission optical microscope with a 60× objective, a halogen lamp as light source, and a spectrometer (Ocean Optics USB2000). Transmission spectra were recorded for bare glass coverslip as reference and for gold plates to obtain the respective transmittance through plates. In order to determine the thickness, these spectra were subsequently fitted in real-time with the multilayer transition model by Tomlin [34] and optical constants for single-crystal gold from [35]. The fitting range is 450 – 740 nm, which reflects the lamp spectrum and the optical feature best. For the fitting only two free parameters were needed: the plate thickness and a flat background. The thickness obtained with this method deviates to the value measured with atomic force microscopy (AFM) by 2 – 5% depending on the thickness and lateral size of the plate.

## 3  Results and Discussion

To illustrate the problems of gold microplates deposited on substrate from dispersion phase, SEM images of microplates drop-casted on a glass coverslip coated with indium tin oxide (ITO) are shown in Figure 1. It is seen that all the large plates in the 1.6-mm-diameter field of view exhibit imperfections. Most imperfections are caused by small gold particles pro-duced during synthesis that attach to the plates. When plates are drop-casted on a substrate, the particles attached to the lower plate surface will create elevations, which extend over a few microns even for submicron particles because of the large stiffness of gold plates. Additionally, the whole area of a plate is often divided into small parts by several elevations. According to literature, the occurrence of small particles during the plate synthesis and the resulting contamination can hardly be avoided. Another problem is plate aggregation, which also causes serious loss of plate area. Furthermore, large microplates show the tendency to bend or even roll up in solution (Figure S2, Supporting Information) due the enormous aspect ratio and probably also due to the accumulation of strain in plates during growth [19,36]. As a consequence, the majority of the microplates prepared on substrate from dispersion phase have defects and a large portion of their area is not suitable for further processing. This is a severe disadvantage especially for fabrication of structures that extend over a large length or area. Moreover, it is also very difficult to thoroughly wash the plates, i.e. to remove synthesis chemicals. Other minor problems such as aging of crystals in solvent have also been reported [19].

In contrast, Figure 2 displays SEM images of typical gold microplates grown on a glass coverslip. To provide the best conditions for SEM, the plates were transferred to a conductive substrate using a PMMA mediated transfer method [37]. Since this transfer method has a nearly 100% yield, the original situation of plate growth on the glass coverslip is represented by Figure 2a. The tilted-view zoom of a plate (Figure 2b) shows that the plate is perfectly attaching to the substrate. Compared to Figure 1, the great improvement in the quality of plates is obvious. Since the plates are growing directly on the substrate, there are no particles beneath the plates, and aggregation or bending does not occur either. Due to the strongly improved plate quality it is now meaningful to further optimize the synthesis protocol to obtain even larger plates, since the complete and continuous plate area can be fully exploited. Substrate-grown gold plates are well suited for post processing, such as washing or surface functionalization, which can be done by just handling the substrate. The aforementioned aging problem is also solved as the plates are stored in dry condition. In addition, Figure 2 demonstrates that by means of transfer, substrate-grown plates can also be used for substrates that are not suitable for plate growth without losing the high quality.

The results of the intermittent growth experiment on glass coverslips (Figure 3) reveal plenty of information about the plate growth. Figure 3a shows a representative overlay of two optical micrographs of the same area of a coverslip before and after regrowth. Obviously, the plates continued to grow in lateral dimensions while maintaining their characteristic regular shapes. This observation of continued growth shows that the plates do grow directly on the substrate and are not formed in solution and then deposited on the substrate. Although the regrowth was based on plates that were already existing on the coverslips, it is reasonable to trace the growth back to earlier stages, e.g. gold atom nucleation or seed formation [19,38,39] that formed on or attached to the coverslips. This is evidenced by the presence of out-of-plane plates (Figure S3, Supporting Information) which must have grown from seeds on substrate [21,40] because it is very unlikely that such large plates form in solution and then attach on a substrate with just one edge or one corner, especially when the substrate is oriented vertically.

The intermittent growth experiment also makes it possible to statistically analyze the growth behavior over time by correlating the two growth periods. Generally the thickness of the plates increased slightly after regrowth (Figure 3b), mostly in the range of 0 – 10 nm. Note that the measurement uncertainty of the thickness measurement is around 5 nm and gets worse for small plates. Therefore some negative increments were obtained for plates with area < 250 µm$^2$ (about 10 µm in length, see Figure S4 in the Supporting Information) before regrowth. Taking the measurement uncertainty into account, we conclude that the thickness increment has no clear dependence on the initial thickness. Moreover, the thickness turned out to be independent of the lateral size as well (Figure S4, Supporting Information).

The correlation between plate areas before and after regrowth ($A_1$ and $A_2$ respectively) is plotted in Figure 3c. It is remarkable that generally there is a clear linear relation between $A_2$ and $A_1$ and the ratio is around 2.5. After some analyses (Supporting Information), we propose the following lateral growth model: the growth of each side facet proceeds such that gold atoms attach to it sequentially (i.e. one atom after the other) and continuously with a certain attachment rate. Such a model is consistent with the screw dislocation model of crystal growth kinetics [41,42] concerning continuous growth and sequential attachment of growth unit. For our case, sequential attachment predicts that the time $\tau$ needed for a side facet to grow one additional complete layer is proportional to the facet's edge length $l$, i.e. $\tau = al$, where $a$ is a facet-dependent factor. Within a time interval d$t$, the number of layers that the side facet grows is d$n$ = d$t$ / $\tau$. We define the growth rate of a side facet as the increment of the number of layers per unit time, $r$ = d$n$ / d$t$ = 1 / (a$l$). The growth rate $r$ is therefore in-versely proportional to the facet length. With each new side facet layer the plate area increases by $\sigma = bl$, where $b$ is a facet-independent factor and describes the pitch of <110> lines along {111} plane. So the area increment within time $dt$ is $dA = \sigma\,dn = \sigma\,(dt/\tau) = (b/a)\,dt$. This equation apparently applies to every side facet of the plate. In the inset of Figure 3c, the histograms of the average area increase rate (area increase $\Delta A$ divided by growth time $\Delta t$) for the initial growth and regrowth period show that both periods have nearly the same distribution of the average area increase rates. This suggests that $a$ is constant in time for every individual side facet. We therefore conclude that the plate area is proportional to the growth time. It is recognized immediately that Figure 3c fits this conclusion very well, as $A_2/A_1 = t_2/t_1 = 3$, where $t_1$ represents the initial growth period and $t_2$ the total growth period (72 hours). The slight deviation of the slope could well be attributed to the growth being slowed down or even stopped before the end of $t_2$ because of reactants depletion. Indeed, the distribution of the average area increase rate for the regrowth period matches that for the initial growth period better if the regrowth time is taken as 1.5 $t_1$ (inset of Figure 3c). The scattering of the data probably results from random variations of the growth rate of the side facets (see below) and variable influence of the growth interruption on the regrowth for each plate.

Another interesting observation is the occurrence of shape transformations of plates during growth. Some examples are depicted in Figure 3d. Obvious shape transformations after regrowth have been observed in less than 1 out of 10 plates and may therefore be considered as rare events. Nevertheless, the study of these shape transformations can contribute to the understanding of growth behavior and mechanism. The crucial feature of the transformed plates is that the side facets have significantly different growth rates. For each of the three plates in Figure 3d, it is clearly seen from the regrowth part (black area) that alternating edges have the same growth rate due to crystal symmetry, while for every two adjacent edges, the growth rates differ a lot. In comparison, for the plates without obvious shape transformation (Figure 3a), the growth rates of every edge are almost the same. An approximate model of the shape evolution is proposed in Figure 3e, showing the schematic of the top {111} plane of a plate. Consider the number of atoms along the plate edge, which represents the length of the edge. One can see that the number of atoms of an edge will decrease or increase by *one* if a new atom layer is formed on the edge itself or one of its neighboring edges, respectively. Thus for edge C in Figure 3e, the atom number can be expressed as $N_{Ct} = N_{C0} + (r_A + r_B - r_C)\,t$, where $N_{C0}$ and $N_{Ct}$ are the atom numbers before and after growth for a time interval $t$, and $r_{A,B,C}$ denote the growth rates of each edge (in most cases $r_A$ equals $r_B$ due to the symmetry of plate). Therefore the length of edge C increases or decreases depending on the sign of $(r_A + r_B - r_C)$. The length variation of all the edges will lead to the change of length ratio between the edges, resulting in shape transformation. Obviously, if the ratio between the growth rates is small (large), the shape evolution will be very slow (fast). Our observation hence reveals that for most of the plates the growth rate difference of side facets is quite small. A special case is the triangle-shaped plates. For a triangular plate, three alternating side facets grow so fast that they just grow out. However, most of the triangular plates retain their shape (for exception see the center panel in Figure 3d) because although the growth of the existing side facets can reconstruct the edges between them, the reconstructed edges will quickly grow out and disappear again.

The variation of lateral size, thickness and shape during growth we have seen so far can be interpreted based on the "seed-based growth" model [19,43]. In the early stage, plate seeds form on the substrate in a totally random way such that each seed has not only a random number of layers of {111} plane, but also a random stacking order of the {111} layers, i.e. a random distribution of twin plans [20,21,44]. The initial thickness of a plate in this picture is determined by the number of {111} layers of the seed. During the following growth stage, gold atoms will add to the top surface and new {111} layers will form, thus increasing the thickness at a certain rate. Since all plates display the same top surface ({111} plane), the increase rate of thickness is also the same for every plate. The stacking order of the {111} layers determines the structures of the six side facets of plates, which are constructions of {111} and {100} planes (Figure S1, Supporting Information) and exhibit different properties in terms of preference for adatoms [20,21,44]. The more preferred by the adatoms, the faster the side facet grows. That is to say, the lateral growth rate is determined by the stacking state of {111} layers of the seed. However, if the facet structure is modified during growth due to stacking of new {111} layers, the growth rate could also change. Since the thickness of a plate is determined by the number of {111} layers of the seed, but has nothing to do with the stacking state, the plate thickness is independent of the lateral dimension. For some plates, the side facets have very different growth rates due to specific stacking structures. Thus the length ratio between faster- and slower-growing side facets is quickly getting smaller and smaller, and shape transformation is therefore observed on these plates easily (left panel of Figure 3d). Furthermore, conversion of faster-growing into slower-growing side facets is also seen in the center and right panels of Figure 3d, which is an indication that the growth rates have changed dramatically. Nevertheless, as these changes are still fully symmetric, they should be attributed to the modification of the stacking structure of {111} layers caused by newly formed layers, as mentioned above.

The intermittent growth experiment on SiO$_2$/Si allows us to have a closer look at the regrown plates with SEM (Figure 4 and Figure S5 in the Supporting Information). The change in the lateral size and shape is very pronounced for the plate in Figure 4. But more importantly, we do not see any trace of the regrowth from the SEM pictures, i.e. there is no boundary between the initial part and regrown part at all. This is also confirmed by the topography image of this plate recorded by AFM (Figure 4c), which does not display any regrowth-related boundary either. This result indicates that the regrowth worked surprisingly well just like the growth had never been interrupted. We therefore propose the possibility of using the regrowth method to increase the plate size. As the lateral growth rate is much higher than the thickness increase rate, we can move the plates to a new solution after the reactants in old solution are depleted to make them regrow and repeat this multiple times. Then the area of the plates will become several times larger but only with very little increase in thickness. Essentially, the surfaces of the regrown plates remain as smooth as without regrowth. Note that simply using larger amount of solution will only result in more plates, instead of larger plates. Therefore, the regrowth method is pretty promising for producing giant plates. Moreover, the traceless regrowth also suggests that there is potential for applying regrowth to nano-structures fabricated out of gold microplates to modify their morphological features.

Apart from glass and SiO$_2$/Si, we also tested other substrates, such as mica, PMMA film and PVA film spin-coated on glass coverslip, ITO coated glass coverslip and Si wafer (type P, orientation <111>, with native oxide). There is no pronounced difference between the former five substrates except that, according to our experience, the microplates grown on PMMA and PVA films tend to be a bit larger than the ones grown on bare coverslip under the same condition. For ITO and Si, however, the situation is quite different. Rather less and smaller microplates were found on ITO compared to the case of bare glass coverslip, while Si was covered by a dense layer of small particles (< 1 μm) with some nanoplates of the same size range among them. This substrate-dependent performance may be attributed to different surface properties of the substrates, for instance, the interfacial free energy between gold and substrate could affect the nucleation and seed formation on the substrate [38,39,45], which has a deterministic effect on the crystal growth. One question of particular interest is whether the electrical conductivity of the substrate (e.g. ITO and Si) has any influence on the plate growth. The impact of a well-defined crystal structure of a crystalline substrate on the plate growth is also worth to be addressed. The study of such substrate-dependent growth behavior may lead to an approach of growing larger plates or controlling the location where plates can form by modifying the substrate surface.

## 4    Conclusion

It has been shown that gold microplates can form and grow directly on substrates immersed in synthesis solution, which proves to be an excellent approach of preparing gold microplates on substrates. Microplates grown on substrate avoid problems suffered by "free" microplates in dispersion phase, and consequently most of the plate area is available for structuring. Direct growth on substrate also allows investigation of the growth behavior of individual plates, because each plate can be tracked easily. Based on intermittent growth experiments, we obtain important insight into the growth mechanism and growth models are proposed. The fact that the plate surface remains smooth after regrowth implies that the regrowth method may be applied to produce giant plates as well as unique single-crystalline nano-structures.

**Acknowledgements**. We thank Jer-Shing Huang, Zhirui Guo, Peter Geisler, Heiko Gross, Andrés Guerrero-Martínez, and Guillerme Stein for stimulating discussions. Financial support by the DFG via HE5618/3-1 (SPP 1391) and HE5618/2-1 is gratefully acknowledged.

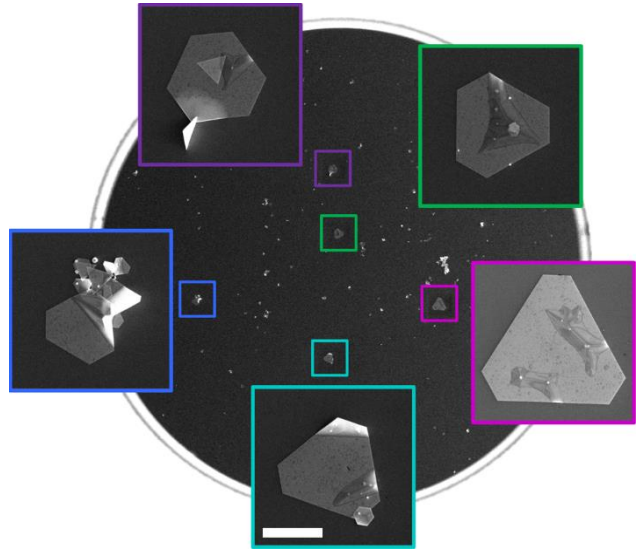

**Figure 1** SEM images of gold microplates drop-casted onto an ITO coated coverslip. The scale bar is 20 µm and applies to all the five zoom images. The diameter of the central overview image is 1.6 mm. The visible small roughness of the plates was caused by the roughness of the ITO coating.

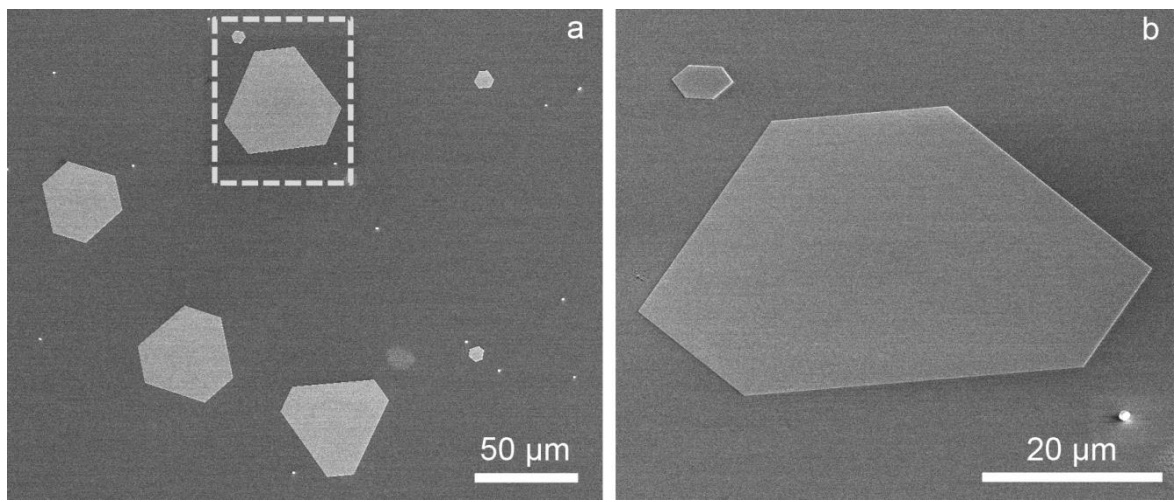

**Figure 2** (a) SEM images of gold microplates grown on a glass coverslip. The plates were transferred to an electrically conductive substrate (coverslip evaporated with gold and SiO$_2$ layers sequentially) for SEM. (b) Zoom to the plate in the dashed rectangle in (a) with a 52° tilt angle.

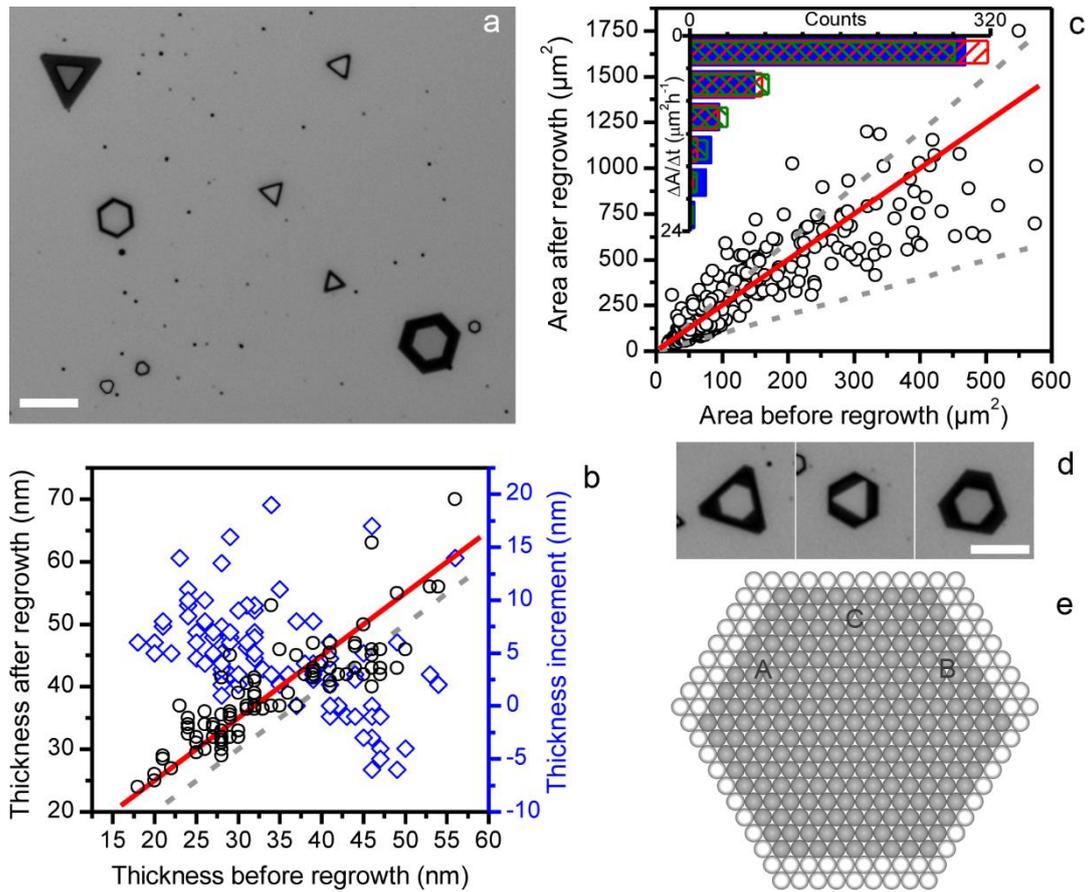

**Figure 3** Regrowth of gold plate on glass coverslip. (a) Overlay of optical microscopy pictures of the same area of a coverslip before and after regrowth. Plate images before and after regrowth are displayed as gray and black respectively. Scale bar, 30 μm. (b) Thickness (black circle) and thickness increment (blue diamond) of individual gold plates after regrowth with respect to their thickness before regrowth. The straight lines are guide to the eye for the black circles with slope of 1 and intercepts of 5 nm (red solid) and 0 (gray dashed), indicating 5 nm and zero thickness increment respectively. (c) Area of individual gold plates after regrowth with respect to their area before regrowth. The straight lines are guide to the eye with slopes of 2.5 (red solid), 3 (upper gray dashed) and 1 (lower gray dashed), respectively. Inset: histograms of the average area increase rate (area increase Δ$A$ over growth time Δ$t$) for the initial growth period with Δ$t$ = 24 h (blue) and regrowth period with Δ$t$ = 48 (red) and 36 h (olive), respectively. (d) Same as (a), but for several plates showing shape transformation. Scale bar, 20 μm. (e) Schematic of the gold atom arrangement in the top {111} plane of a gold plate. The hollow circles denote the new adatoms on the six side facets, demonstrating the effect of growth rate differences.

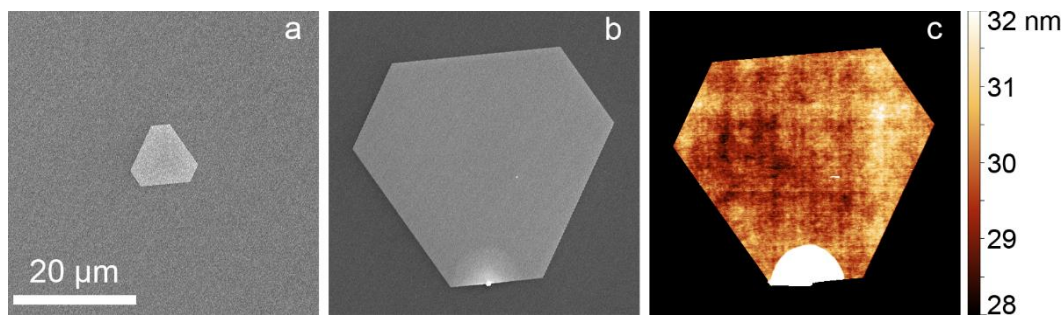

**Figure 4** Regrowth of gold plate on SiO$_2$/Si substrate. (a) and (b) SEM images of a gold plate before and after regrowth respectively. The real relative orientation between the two plate images is represented. (c) AFM topography image of the plate after regrowth. The scale bar applies to all the three panels.

# Supporting Information

- **Schematic gold plate model and structures of side facets**

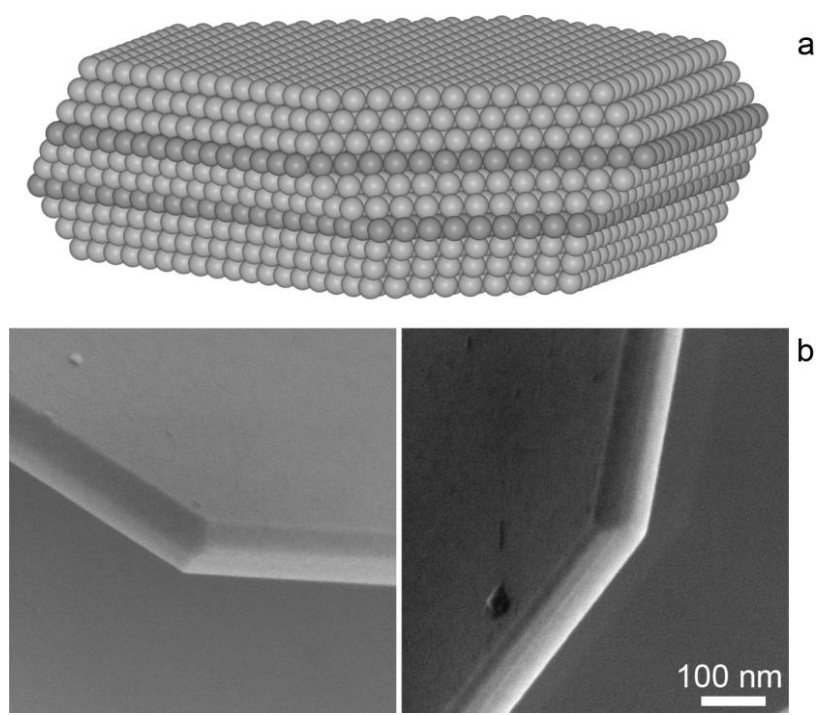

**Figure S1.** (a) Schematic gold plate model constructed with spheres with two twin planes (slightly darker layers). (b) SEM pictures of side facets of gold plates.

It is seen from the gold plate model in Figure S1a that the side facets are constructions of {111} and {100} planes. Specific structures of side facets of real gold microplates are shown in the SEM pictures in Figure S1b.

- **Bending gold microplates**

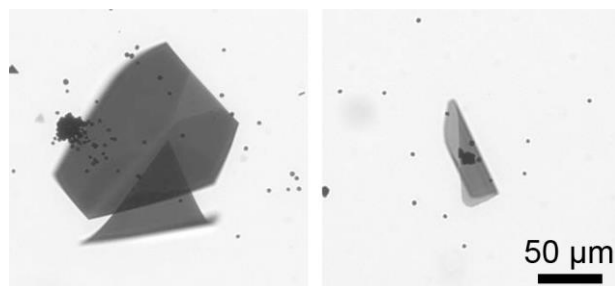

**Figure S2.** Optical micrographs of bending gold microplates.

Several gold microplates that are bending or rolling up are shown in Figure S2. The substrate is a coverslip that was immersed in synthesis solution for plate growth. After the synthesis, the coverslip was taken out of the solution and then directly put on the microscope for observation without further processing. Therefore there was a layer of synthesis solution on the coverslip and some plates in the solution layer were brought to the coverslip, which were recognized if they moved along with the solution on the coverslip when the coverslip was slightly tilted. The plates in Figure S2 were found in the solution layer. In contrast, the plates grown on the same coverslip with similar dimensions (lateral size and thickness) were found attaching on the coverslip flatly. No stirring or strong shaking was applied to the synthesis solution.

- **Out-of-plane gold microplates**

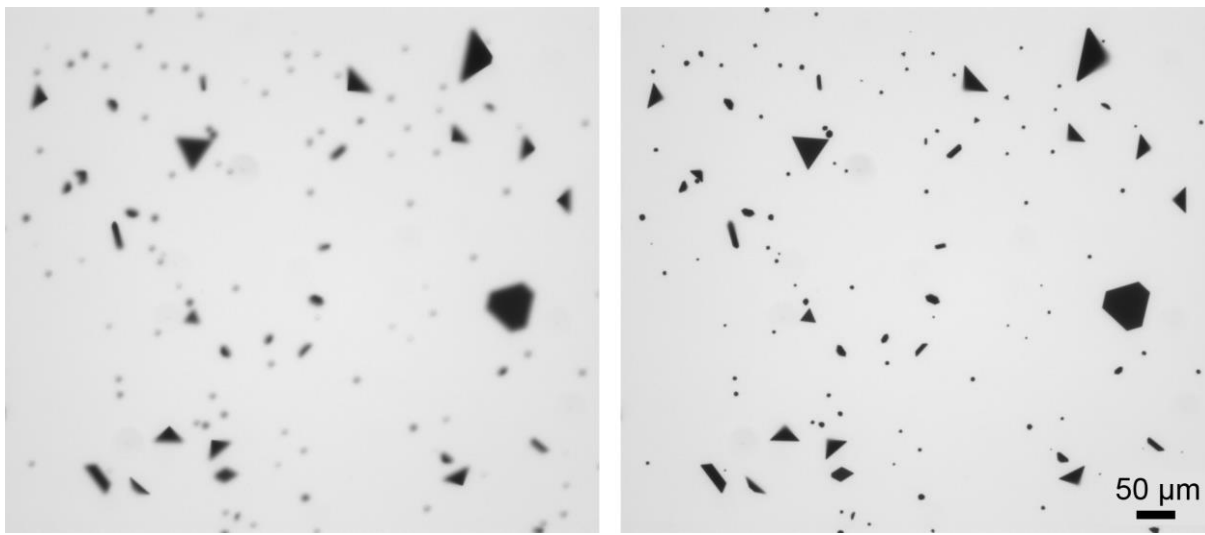

**Figure S3.** Gold microplates growing out-of-plane on glass coverslips imaged by transmission optical microscopy. The two pictures show the same area of a coverslip with slightly changed focus. Several out-of-plane microplates can be identified.

The observation condition for Figure S3 is the same as that for Figure S2. After plate growth, the coverslip was taken out of the synthesis solution and directly put on the microscope without further processing. Hence there was a layer of solution on the coverslip as well. From Figure S3, it is clearly seen that some microplates are standing on the surface with a certain angle, as their two opposite parts are not in focus at the same time. These out-of-plane microplates are also often seen when the substrate is still in solution. The reflection with arbitrary angles by the plates on the substrate and the vibration of the plates when slightly shaking the solution verify that these plates are attached on the substrate but pointing out of the surface with arbitrary orientation. It should be noticed that the standing plates are not seen in Figure 2 and 3 because they were washed away since their bonding with the substrate is very weak.

- **Thickness versus area**

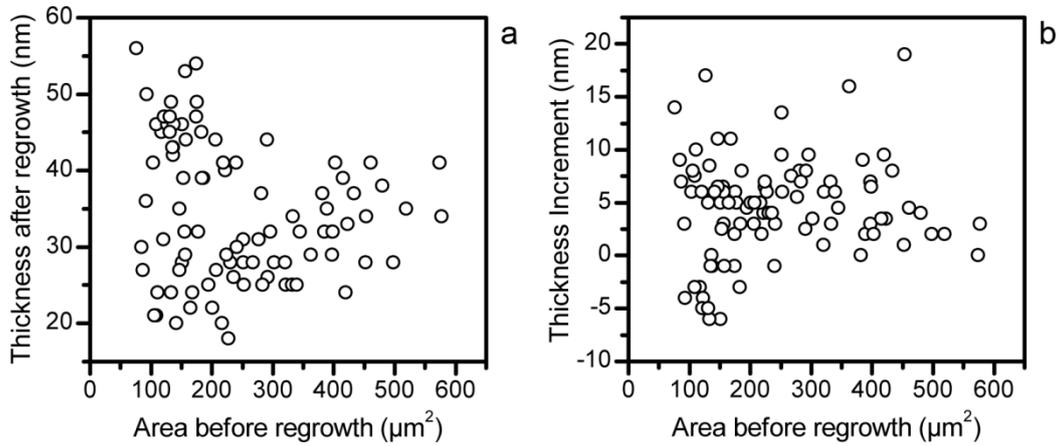

**Figure S4.** Thickness (a) and thickness increment (b) of individual gold plates after regrowth with respect to their area before regrowth.

The two plots in Figure S4 indicate that the plate thickness is independent of the lateral dimension.

- **Lateral growth model of gold microplates**

In contrary to the model proposed in the main text, where gold adatoms attach to side facets sequentially and the growth rate is a function of edge length, another possible model is supposing that gold adatoms attach to side facets and form every segment along the edges simultaneously. Consequently, in this model the growth rate of a side facet (same definition as in main text) is independent of the edge length. Considering a triangular or hexagonal plate of which all side facets have the same growth rate $r$, it is easily derived that its area $\propto (rt)^2$, where $t$ is the growth time. $r$ is regarded as a constant over time because the only factor that could induce the change of growth rate is the variation of the side facet structure that results from the thickness increase. In general this type of growth rate change is negligible as the relative thickness change is quite small (see main text). Consequently, we get $A \propto (t)^2$, which means $A_2/A_1 = (t_2/t_1)^2 = 9$, apparently different from Figure 3c. As a conclusion, this simultaneous attaching model does not fit the experimental results.

The jump of growth rate due to the growth interruption by the measurement before regrowth can also be excluded from the possible interpretation of Figure 3c, because this kind of influence would be random rather than leading to the collective behavior.

- **Regrowth of gold microplates on SiO₂/Si substrate**

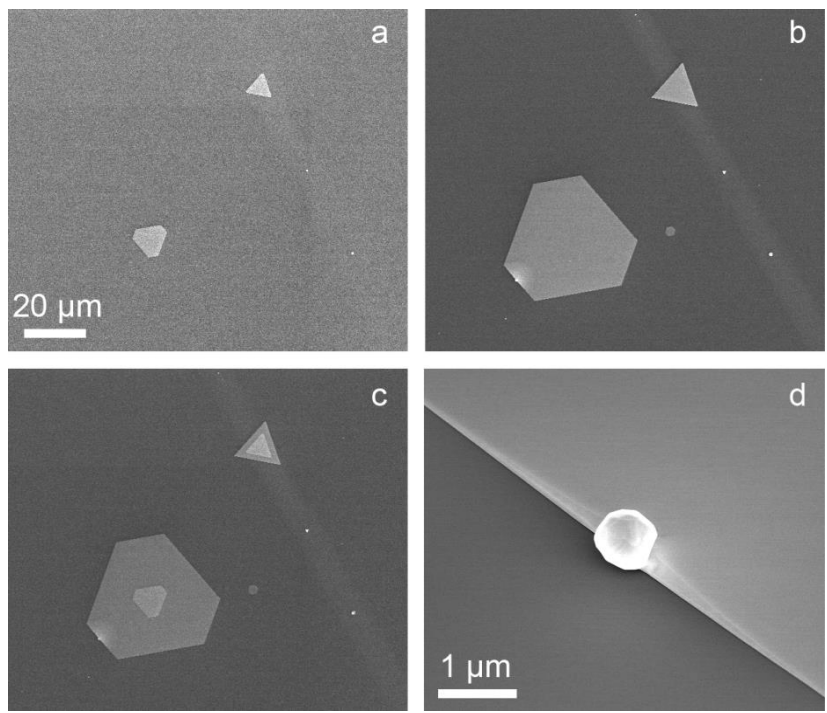

**Figure S5.** SEM images of gold plates before (a) and after (b) regrowth on SiO$_2$/Si substrate, and their overlay (c). The straight line in the images is a part of the maker pattern which was milled on the substrate using FIB. The zoom of a bump and a particle at the lower-left edge of the bigger plate in (b) is shown in (d), with a 52° tilt angle.

In addition to Figure 4, Figure S5a-c show more details about how the SEM images of gold plates before and after regrowth are compared. In Figure S5d, the bump and particle are seen to be on the top side of the plate, probably because the particle attached to the edge from the solution and affected the growth of the edge, and then the bump started to form there as a result.